\documentclass{webofc}
\usepackage[varg]{txfonts}

\usepackage{hyperref}
\usepackage{listings}
\usepackage{caption}
\usepackage{subcaption}
\usepackage[export]{adjustbox}
\begin{document}
\title{Experience deploying an analysis facility for the Rubin Observatory’s Legacy Survey of Space and Time (LSST) data}

\author{
\firstname{Gabriele} \lastname{Mainetti}\inst{1}\fnsep\thanks{\email{gabriele.mainetti@cc.in2p3.fr}} \and
\firstname{Fabio} \lastname{Hernandez}\inst{1}
\and
\firstname{Fabrice} \lastname{Jammes}\inst{2}
\and
\firstname{Quentin} \lastname{Le Boulc'h}\inst{1}
}

\institute{
    CNRS, CC-IN2P3, 21 avenue Pierre de Coubertin, CS70202, 69627 Villeurbanne CEDEX, France 
    \and
    Université Clermont Auvergne, CNRS/IN2P3, Laboratoire de Physique de Clermont, F-63000 Clermont-Ferrand, France
}

\abstract{%
  The Vera C. Rubin Observatory is preparing for the execution of the most ambitious astronomical survey ever attempted, the Legacy Survey of Space and Time (LSST). Currently in its final phase of construction in the Andes mountains in Chile and due to start operations  in 2025 for 10 years, its 8.4-meter telescope will nightly scan the southern sky and collect images of the entire visible sky every 4 nights using a 3.2 Gigapixel camera, the largest imaging device ever built for astronomy. Automated detection and classification of celestial objects will be performed by sophisticated algorithms on high-resolution images to progressively produce an astronomical catalog eventually composed of 20 billion galaxies and 17 billion stars and their associated physical properties.

In this paper, we briefly present the infrastructure deployed at the French Rubin data facility (operated by IN2P3 computing center, CC-IN2P3) to deploy the Rubin Science Platform, a set of web-based services to provide effective and convenient access to LSST data for scientific analysis. We describe the main services of the platform, the components that provide those services and our deployment model. We also present the Kubernetes-based infrastructure we are experimenting with for hosting the LSST astronomical catalog, a petabyte-scale relational database developed for the specific needs of the project.
}
\maketitle
\section{Introduction}
\label{intro}
This article outlines the tools being deployed by the CNRS/IN2P3 Computing Center\footnote{https://cc.in2p3.fr} (\href{https://cc.in2p3.fr}{CC-IN2P3}) to provide researchers a convenient way to explore and analyse the data that will be produced as part of the Legacy Survey of Space and Time (LSST) by the \href{https://rubinobservatory.org}{Vera C. Rubin Observatory}.
 
The Vera C. Rubin Observatory is in the final stages of its construction on the Chilean Andes. Equipped with an 8.4-meter telescope and the largest digital camera ever built (3.2 gigapixel), it will execute the most ambitious survey ever attempted: the Legacy Survey of Space and Time (LSST) collecting imagery covering the entire visible sky every four nights. 

The four main science goals of LSST are to investigate dark matter and dark energy, map the Milky Way, take an inventory of the solar system and study the transient sky\,\cite{Ivezic:2019}. To achieve those goals, the Observatory will produce, by the end of the survey after 10 years of operation, an astronomical catalog containing 37 billion objects (20 billion galaxies, 17 billion stars) as well as a set of 5.5 million images. Science-ready images and the astronomical catalog will be regularly delivered by the Observatory to science collaborations.

This paper is structured as follows. The section \ref{ccin2p3} introduces the role of CC-IN2P3 as a data facility for the Rubin Observatory. Section \ref{analysis_intro} describes the analysis facility, the technology behind it and the resources deployed by the CC-IN2P3 to run it. The following sections cover the analysis facility main components and, in particular, section \ref{qserv} describes the database component and section \ref{rsp} the data analysis platform.

\section{CC-IN2P3 as a Rubin Observatory Data Facility}
\label{ccin2p3}

\begin{figure}[h]
\includegraphics[width=\textwidth]{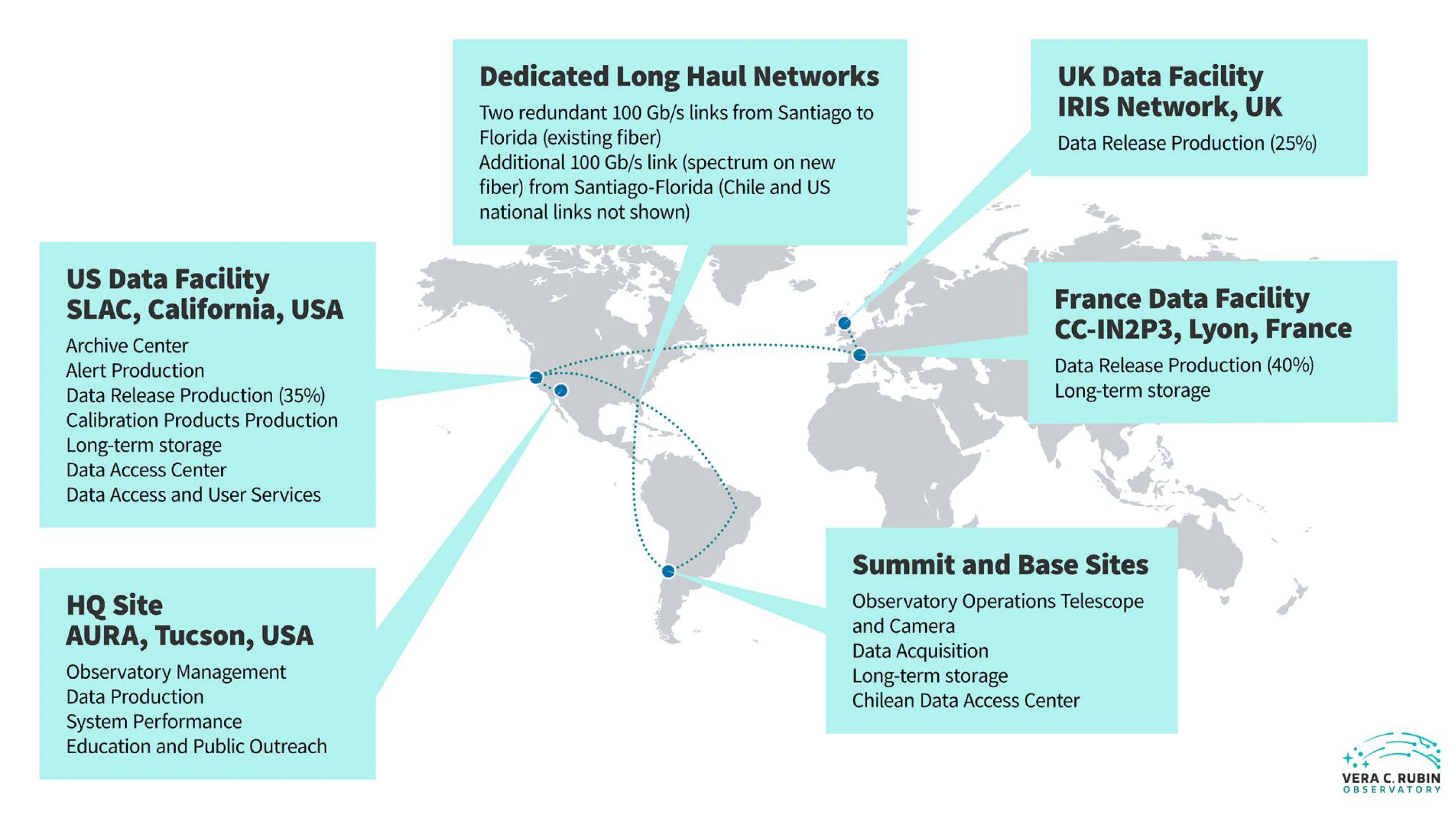}
\caption{Images flow from the Summit Site where the telescope is located in Chile to the three Rubin Data Facilities which collectively provide the computational capacity for processing the images taken by the Observatory for the duration of the survey.}
\label{fig:data-facilities}
\end{figure}

Image data recorded by the Rubin Observatory will be processed by 3 centers, one in the USA (US Data Facility - USDF\footnote{USDF is located at SLAC National Accelerator Laboratory, \href{https://www.slac.stanford.edu}{https://www.slac.stanford.edu}}) and the other two in Europe (UK Data Facility - UKDF\footnote{Supported by the IRIS Infrastructure \href{https://www.lsst.ac.uk}{https://www.lsst.ac.uk} and \href{https://www.iris.ac.uk}{https://www.iris.ac.uk}} and France Data Facility - FrDF), see Figure \ref{fig:data-facilities}.

FrDF is located in Lyon, France at the CC-IN2P3 and will provide the computing and storage resources to annually process 40\% of the raw images recorded since the beginning of the survey, with 35\% provided by the USDF and 25\% by the UKDF.
In addition, CC-IN2P3 is preparing to provide long-term storage for the entire raw images data set as well as for selected data products. By the end of the survey, the size of the Rubin data sets could reach several hundred petabytes, with 15 PB for the final version of the astronomical catalog alone.

\section{The analysis facility}
\label{analysis_intro}

The Rubin Observatory's data analysis facility deployed at CC-IN2P3 is designed with three main objectives: to provide access to Rubin data, to integrate the working environment with the rest of the site's environment and to reach high levels of scalability and resilience.

\subsection{Data Access}
Our main objective is to provide researchers with an intuitive platform providing a convenient and effective way to access and to analyse  the extensive survey data (both images and catalog) generated by the Rubin Observatory. This objective is achieved with two major components of the platform: 
 
 \begin{enumerate}
      
 \item \textbf{Qserv}: Qserv\,\cite{qserv, qservrepo} is an open source Massively Parallel Processing database which hosts the Rubin astronomical catalog (see Section \ref{qserv} below for details), and
\item \textbf{Rubin Science Platform (RSP)}: the RSP\,\cite{LSE-319} is an online interactive analysis hub, allowing researchers to access and analyse the data via a user-friendly web-based interface, see Section \ref{rsp} for more details. 
 \end{enumerate}

\subsection{Integration with CC-IN2P3 Environment}

We want the analysis facility to be seamlessly integrated into the existing CC-IN2P3 working environment to make easy for the researchers the transition between the platform and their familiar environment. This integration involves many aspects, including authentication and transparent access to all the storage areas individual users can usually access from a terminal session or from the batch farm, including their \texttt{\$HOME} directory and other project-specific disk-based storage areas. Site-wide single sign on authentication at CC-IN2P3 is based on Keycloak\,\cite{keycloak}.

\subsection{Scalability and Resilience}

The platform is designed to be both scalable and resilient to ensure that it can adapt to evolving demands linked to the large volumes of astronomical data. The use of Kubernetes (K8S)\,\cite{k8s} as the orchestration tool for the platform's containers allows us to effectively manage and scale the components as needed.

For this purpose, CC-IN2P3 provides two K8S clusters dedicated to the Qserv and the Rubin Science Platform: one used as test-bench is based on OpenStack\,\cite{openstack} virtual machines and another one used for production on top of bare-metal\footnote{The bare metal cluster is composed of DELL PowerEdge R440 and DELL PowerEdge R540 servers}, see Table \ref{cluster_table} for a summary.

\begin{table}[h]
\centering
\caption{Details of the components of the Kubernetes clusters dedicated to the Rubin analysis platform at CC-IN2P3.}
\label{cluster_table}       % Give a unique label
% For LaTeX tables you can use
\begin{tabular}{l|cc}
\hline
Cluster & Production & Test-bench  \\\hline
Infrastructure & Bare-metal & OpenStack \\
Total Nodes & 25 & 8 \\
Qserv Nodes & 17 & 3 \\
RSP Nodes & 5 & 1 \\
Local Storage & 50 TB & 1 TB \\\hline
\end{tabular}
\end{table}

CC-IN2P3 provides also 4 data transfer nodes to share with the community catalog data ready for ingestion into Qserv via Caddy\,\cite{caddy} HTTP servers; these servers are also used internally to populate the Qserv databases deployed at the site (see Section \ref{ingest}). 

\section{Qserv}
\label{qserv}
The astronomical catalog produced by LSST will include the physical properties of 20 billion galaxies and 17 billion stars, resulting in a total of 15 petabytes of data by the end of the survey. The Qserv database management system \cite{qserv, qservrepo} has been specifically developed by Rubin project members at SLAC, with contributions from IN2P3, to handle this large volume of data and to respond to the specific data extraction requests scientists typically perform against astronomical catalogs.

\begin{figure}[h]
\includegraphics[width=0.7\textwidth, center]{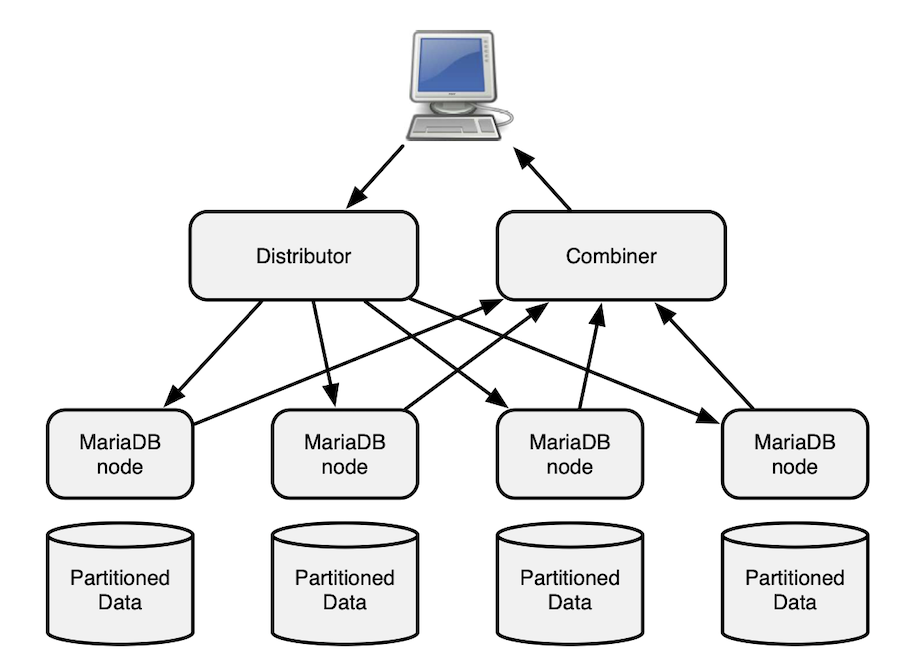}
\caption{Shared-nothing database architecture, from \cite{DMTN-243}}
\label{fig:qserv_schema}
\end{figure}

Qserv is a shared-nothing Massively Parallel Processing Relational (SQL) Database: this means the processing is split among servers and a leader node handles communications with each of the individual nodes as a map-reduce process. Nodes don't share resources with other nodes. A representation of this architecture is shown in Figure \ref{fig:qserv_schema}.

Developed as an open source software project, it includes data partitioning over the celestial sphere's region of equal area, data replication to ensure resilience and high-availability, shared scanning to reduce the total I/O load, and sciSQL\,\cite{scisql} User Defined Functions (UDFs) to simplify spherical geometry, statistics and photometry queries. 

\subsection{Deployment: \texttt{qserv-operator}}
Deployment of Qserv is based on Kubernetes operator SDK\,\cite{operator}, a framework to manage and automate the deployment of complex applications. This approach makes the installation of Qserv really easy because only two commands are necessary\footnote{A demo of the deployment is available at \href{https://is.gd/FK62Wa}{https://is.gd/FK62Wa}.}: 

\begin{verbatim}
  $ kubectl apply -f manifest/operator.yaml
  $ kubectl apply -k manifest/<instance>
\end{verbatim}

\noindent where \texttt{<instance>} refers to the customization needed for a specific infrastructure.  The \texttt{qserv-operator}\,\cite{qserv-operator} code is available in \href{https://github.com/lsst/qserv-operator}{GitHub}.

\subsection{Data ingestion: \texttt{qserv-ingest}}
\label{ingest}

Qserv has a powerful distributed data ingestion algorithm integrated: \texttt{qserv-ingest}. 

\begin{figure}[h]
\includegraphics[width=\textwidth, center]{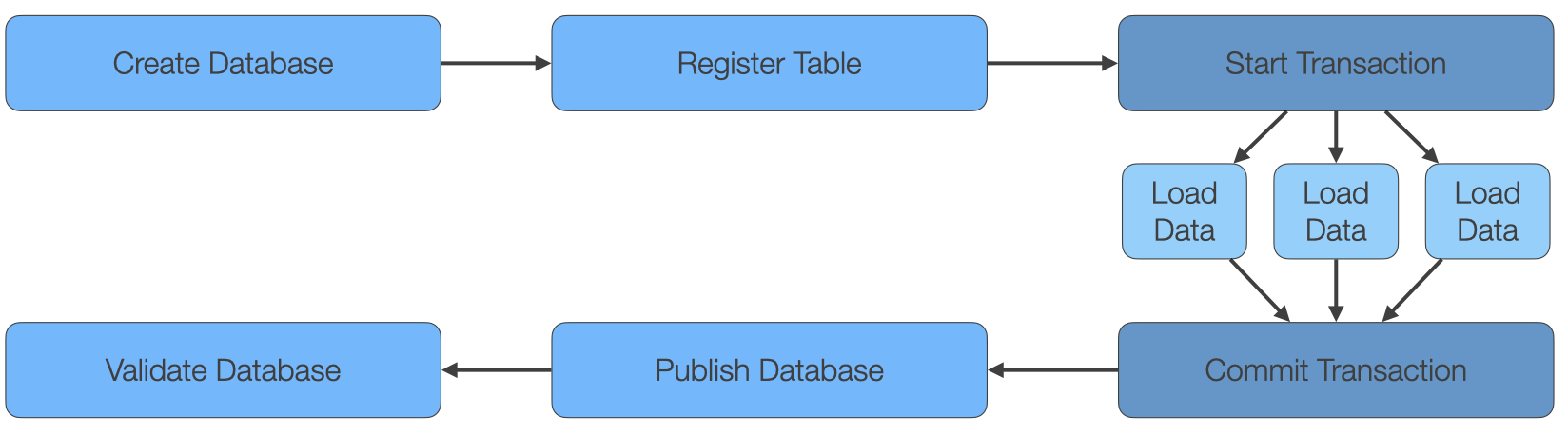}
\caption{Schematic view of the data ingestion process into Qserv database.}
\label{fig:ingest_schema}
\end{figure}

The ingestion workflow's schema is shown in Figure \ref{fig:ingest_schema}; \texttt{qserv-ingest} uses Argo Workflow\,\cite{argowf} to orchestrate the ingestion tasks which concurrently execute on all nodes in the cluster. The workflow's input data to be loaded into the catalog's tables must be previously partitioned (using a partitioner included in Qserv) and stored in CSV format. 

Each transaction can handle a significant number of files for ingestion. Asynchronous REST requests are issued to drive the ingestion of each input file, enabling recovery in case of errors. The workflow also performs validation steps and executes benchmarks on the ingested database. 

During a test campaign in 2022, \texttt{qserv-ingest} was able to ingest 22 million files for an aggregated 40TB of data in 5 hours. At CC-IN2P3 we use \texttt{qserv-ingest} as the workflow to ingest data in Qserv. 

\subsection{Status of Qserv at CC-IN2P3}
\label{qserv_status}

The production Qserv instance at CC-IN2P3 is currently composed of 17 nodes, 15 qserv-worker and 2 qserv-master nodes. There are 5 astronomical catalogs ingested for an aggregated size of 90 TB, as shown in Table  \ref{qserv_table}. 

\begin{table}[h]
\centering
\caption{Status of Qserv-managed astronomical catalogs at CC-IN2P3.}
\label{qserv_table}       % Give a unique label
% For LaTeX tables you can use
\begin{tabular}{l|cc}
\hline
Catalog & Size (TB) & Number of rows (Billions) \\\hline
idf-dp0.2-catalog & 36.6 & 139 \\
frdf-dp0.2-catalog & 35.9 & 121 \\
dp01\_dc2\_catalogs & 1.1 & 1.7 \\
skysim5000\_v1.1.1 & 13.6 & 20.5 \\
cosmoDC2\_v1.1.4 & 3.7 & 5.5  \\\hline
\end{tabular}
\end{table}

The DP0.1 and DP0.2 catalogs above are catalogs generated by the LSST data processing campaigns\,\cite{RTN-001, pipecc} using simulated sky images and they are a preview of what we expect once the Rubin Observatory will start to produce real data.

\section{The Rubin Science Platform}
\label{rsp}

The \textbf{Rubin Science Platform} (RSP) is a coherent set of web-based applications and services designed by the Rubin project's Data Management team to offer scientists a unified, near-to-the-data interactive analysis tool. It uses Firefly\,\cite{firefly} for data visualisation and plotting and provides access to both Qserv-managed catalogs as well as external catalogs. For advanced, programmatic analysis using Python it integrates a Jupyter\,\cite{jupyter} notebook platform configured to include the LSST Science Pipelines\,\cite{lsst-science-pipelines}. Finally, it acts also as gateway to Qserv catalogs for Virtual Observatory\footnote{https://www.ivoa.net} Table Access Protocol\,\cite{TAP} compatible tools. 

\begin{figure*}
\centering
    \begin{subfigure}[b]{\textwidth}
         \centering
         \includegraphics[width=\textwidth]{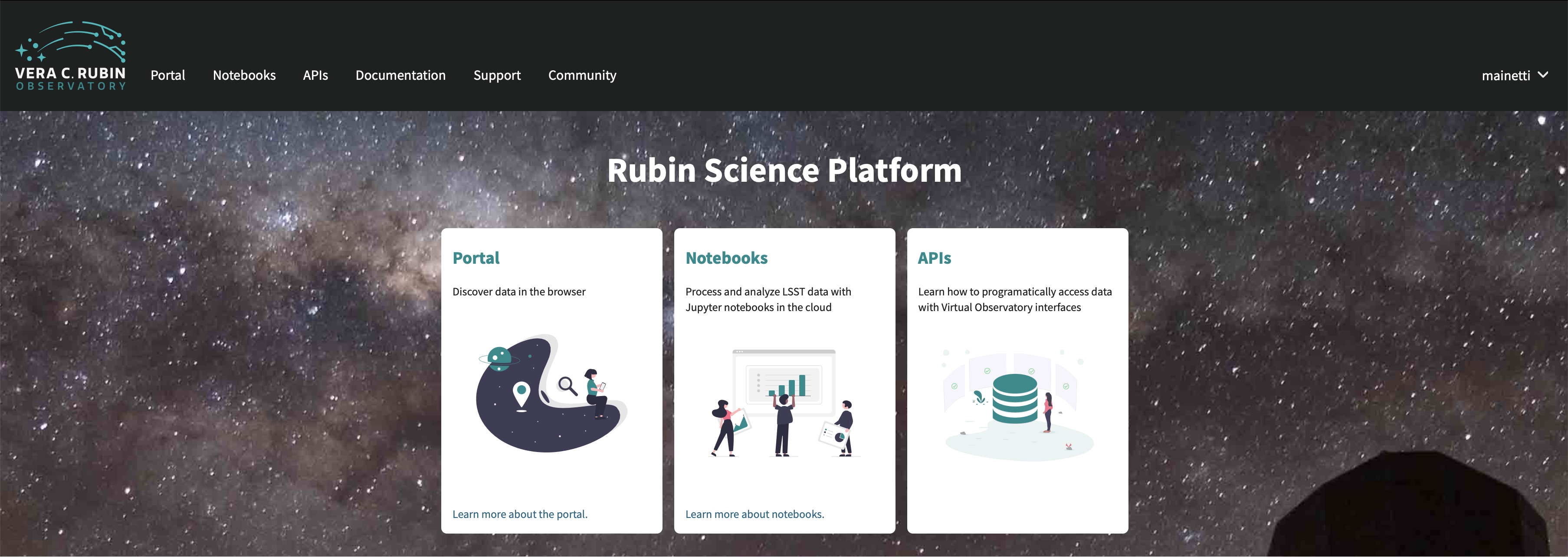}
         \caption{Web portal access}
         \label{fig:portal}
     \end{subfigure}
    \begin{subfigure}[b]{0.45\textwidth}
         \centering
         \includegraphics[width=\textwidth]{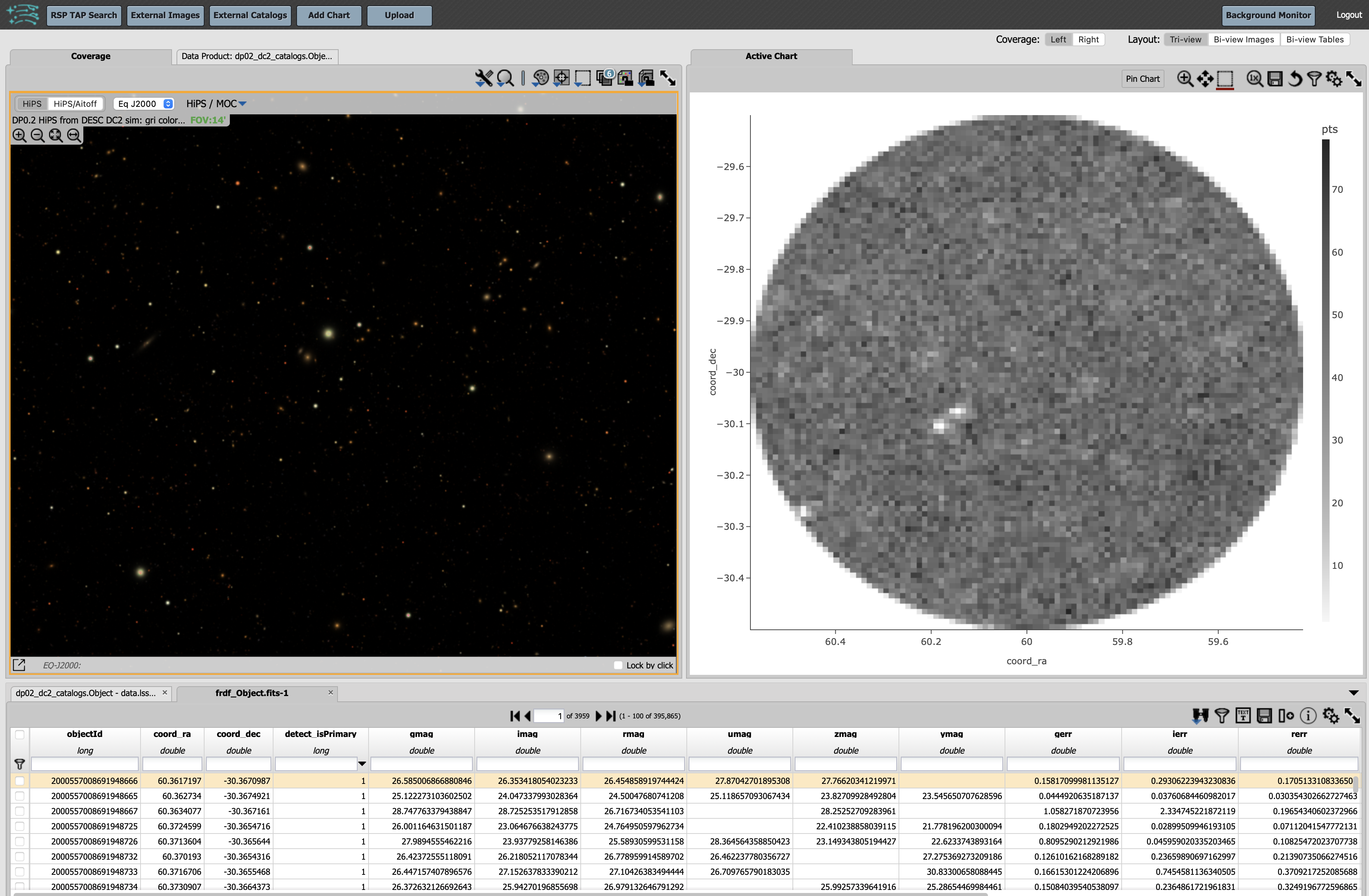}
         \caption{Tables and images viewer}
         \label{fig:RSP_DW}
     \end{subfigure}
     \hfill
     \begin{subfigure}[b]{0.45\textwidth}
         \centering
         \includegraphics[width=\textwidth]{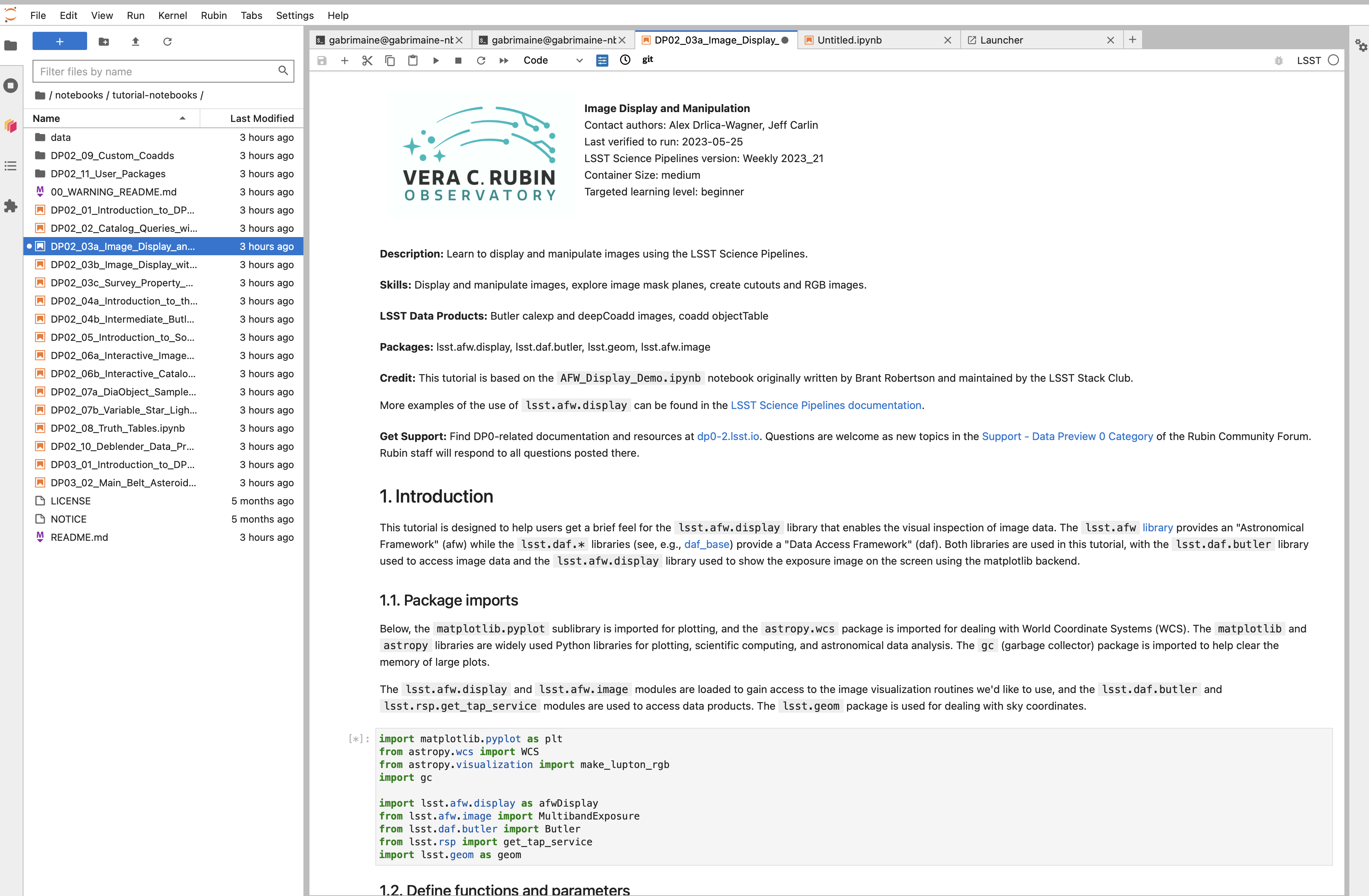}
         \caption{Jupyter notebook platform}
         \label{fig:JNP}
     \end{subfigure}
     % Give the correct figure height in cm
\caption{A view of the RSP}
\label{fig:rsp}       % Give a unique label
\end{figure*}

A view of the RSP is shown in Figure \ref{fig:rsp}.

An experimental instance of the Rubin Science Platform deployed at CC-IN2P3 is online and accessible to registered project members at the address \href{http://data-dev.lsst.eu}{http://data-dev.lsst.eu}.

\subsection{Deployment}

Designed as a collection of applications and services, it has been developed by the Rubin Observatory Science Quality and Reliability Engineering\footnote{https://github.com/lsst-sqre} (SQuaRE) team.

The RSP components run on top of Kubernetes clusters and each component is configured via Helm \cite{helm} charts. Individual Helm charts are managed via the Phalanx\,\cite{phalanx} git repository\,\cite{phalanxrepo}.

The use of ArgoCD\,\cite{argocd} allows to synchronize these application deployment manifests into the Kubernetes cluster of each environment.

Authentication is managed via the \textbf{gafaelfawr} (\cite{gafaelfawr}, \cite{DMTN-224}, \cite{DMTN-234}) identity management application, which supports LDAP\,\cite{ldap} and OpenId Connect\,\cite{opeind} providers (e.g. Github). CC-IN2P3's RSP instance uses Keycloak\,\cite{keycloak} as OpenID Connect (OIDC) provider and LDAP as provider of the user's group identifiers.

In addition to gafaelfawr, there are 4 core applications providing key functionality for other applications:

 \begin{itemize}
    \item \textbf{argocd} for deployment orchestration,
    \item \textbf{cert-manager} for certificates management,
    \item \textbf{ingress-nginx} for traffic routing,
    \item \textbf{vault-secrets-operator} for secret management
 \end{itemize}

Seventeen components are currently activated on the CC-IN2P3 RSP instance, some of them requiring adaptations to the specifics of the CC-IN2P3 environment.    

\section{Summary}

The Rubin Observatory will record a large number of images of the Southern sky over several years. We outlined the role of CC-IN2P3 as one of the three main data facilities that will process these images to regularly produce an updated version of an astronomical catalog and we presented how we deployed an analysis facility on top of Kubernetes, highlighting its two key components: the scalable, shared and resilient database designed for serving the Rubin astronomical catalog and the Rubin Science Platform to perform interactive analysis of the data to be regularly released by the Observatory.

\bibliography{references}

\begin{thebibliography}{28}

\bibitem{Ivezic:2019}
Z.~Ivezic et~al., Astrophys. J. \textbf{873}, 111 (2019)

\bibitem{qserv}
D.L. Wang et~al., \emph{Qserv: A Distributed Shared-Nothing Database for the
  LSST Catalog}, in \emph{State of the Practice Reports} (Association for
  Computing Machinery, 2011), SC '11, ISBN 9781450311397,
  \urlstyle{tt}\url{https://doi.org/fnk9nm}

\bibitem{qservrepo}
\emph{{Qserv}: petascale distributed database},
  \url{https://github.com/lsst/qserv}

\bibitem{LSE-319}
{M. Jurić, D. Ciardi ,G.P. Dubois-Felsmann and L.P. Guy}, \emph{{LSE-319 -
  LSST Science Platform Vision Document}} (2019), {Vera C. Rubin Observatory
  Systems Engineering Technical Note},
  \urlstyle{tt}\url{https://lse-319.lsst.io}

\bibitem{keycloak}
\emph{{KeyCloak}}, \url{https://www.keycloak.org}

\bibitem{k8s}
\emph{{Kubernetes}}, \url{https://kubernetes.io}

\bibitem{openstack}
\emph{{OpenStack}}, \url{https://www.openstack.org}

\bibitem{caddy}
\emph{{Caddy}, an open source web server}, \url{https://caddyserver.com}

\bibitem{DMTN-243}
F.~Mueller et~al., \emph{{DMTN-243 - Qserv: A Distributed Petascale Database
  for the LSST Catalogs}} (2022), {Vera C. Rubin Observatory Data Management
  Technical Note}, \urlstyle{tt}\url{https://dmtn-243. lsst.io/}

\bibitem{scisql}
\emph{{sciSQL 0.3}: Science tools for mysql},
  \url{https://smonkewitz.github.io/scisql/}

\bibitem{operator}
\emph{{Operator Framework} for kubernetes}, \url{https://operatorframework.io}

\bibitem{qserv-operator}
\emph{{qserv-operator: a qserv operator for Kubernetes}},
  \url{https://github.com/lsst/qserv-operator}

\bibitem{argowf}
\emph{{Argo Workflow} a workflow engine for orchestrating parallel jobs on
  kubernetes}, \url{https://argoproj.github.io/argo-workflows}

\bibitem{RTN-001}
W.~O'Mullane, \emph{{RTN-001 - Data Preview 0: Definition and planning.}}
  (2021), {Vera C. Rubin Observatory Technical Note},
  \urlstyle{tt}\url{https://rtn-001.lsst.io/}

\bibitem{pipecc}
Q.~{Le Boulc'h}, F.~Hernandez, G.~Mainetti, \emph{{The Rubin Observatory’s
  Legacy Survey of Space and Time DP0.2 processing campaign at CC-IN2P3}}, in
  \emph{{Proc. of CHEP 2023}} ({to appear})

\bibitem{firefly}
\emph{{Firefly} a browser-based interactive particle visualization app},
  \url{http://firefly-viz.com}

\bibitem{jupyter}
\emph{Jupyter}, \url{https://jupyter.org}

\bibitem{lsst-science-pipelines}
J.~{Bosch} et~al., \emph{{An Overview of the LSST Image Processing Pipelines}},
  in \emph{Astronomical Data Analysis Software and Systems XXVII}, edited by
  P.J. {Teuben}, M.W. {Pound}, B.A. {Thomas}, E.M. {Warner} (2019), Vol. 523 of
  \emph{Astronomical Society of the Pacific Conference Series}, p. 521,
  \texttt{1812.03248}

\bibitem{TAP}
P.~{Dowler}, G.~{Rixon}, D.~{Tody}, M.~{Demleitner}, \emph{{Table Access
  Protocol Version 1.1}}, IVOA Recommendation 27 September 2019 (2019),
  \urlstyle{tt}\url{https://10.5479/ADS/bib/2019ivoa.spec.0927D}

\bibitem{helm}
\emph{{Helm} the package manager for kubernetes}, \url{https://helm.sh}

\bibitem{phalanx}
\emph{{Phalanx} a gitops repository for rubin observatory’s kubernetes
  environments}, \url{https://phalanx.lsst.io}

\bibitem{phalanxrepo}
\emph{{Phalanx}}, \url{https://github.com/lsst-sqre/phalanx}

\bibitem{argocd}
\emph{{ArgoCD} a declarative, gitops continuous delivery tool for kubernetes},
  \url{https://argo-cd.readthedocs.io}

\bibitem{gafaelfawr}
\emph{{Gafaelfawr} rsp identity management},
  \url{https://github.com/lsst-sqre/gafaelfawr}

\bibitem{DMTN-224}
{R. Allbery}, \emph{{DMTN-224 - RSP identity management implementation
  strategy}} (2023), {Vera C. Rubin Observatory Data Management Technical
  Note}, \urlstyle{tt}\url{https://dmtn-224.lsst.io/}

\bibitem{DMTN-234}
{R. Allbery}, \emph{{DMTN-234 - RSP identity management design}} (2023), {Vera
  C. Rubin Observatory Data Management Technical Note},
  \urlstyle{tt}\url{https://dmtn-234.lsst.io/}

\bibitem{ldap}
\emph{{LDAP}}, \url{https://ldap.com}

\bibitem{opeind}
\emph{{OpenId}}, \url{https://openid.net}

\end{thebibliography}
\end{document}